# Characterizing human collective behaviours of COVID-19 in Hong Kong

Zhanwei Du[1+], Xiao Zhang[2+], Lin Wang[3+], Sidan Yao[4+], Yuan Bai[1,2], Qi Tan[1,2], Xiaoke Xu[5], Sen Pei[6], Jingyi Xiao[1], Tim K. Tsang[1], Qiuyan Liao[1], Eric H. Y. Lau[1,2], Peng Wu[1,2], Chao Gao[7*], Benjamin J. Cowling[1,2*]

[1] WHO Collaborating Centre for Infectious Disease Epidemiology and Control, School of Public Health, Li Ka Shing Faculty of Medicine, The University of Hong Kong, Hong Kong Special Administrative Region, China.

[2] Laboratory of Data Discovery for Health, Hong Kong Science and Technology Park, New Territories, Hong Kong Special Administrative Region, China

[3] Department of Genetics, University of Cambridge, Cambridge, CB2 3EH, UK

[4] Institute of High Performance Computing (IHPC), Agency for Science, Technology and Research (A*STAR), Singapore

[5] College of Information and Communication Engineering, Dalian Minzu University, Dalian, China

[6] Department of Environmental Health Sciences, Mailman School of Public Health, Columbia University, New York, NY 10032, USA

[7] School of Artificial Intelligence, Optics, and Electronics (iOpen), Northwestern Polytechnical University, Xi'an, China

[+] These authors are joint first authors with equal contribution.
[*] Corresponding author: Benjamin J. Cowling and Chao Gao

**Abstract**

People are likely to engage in collective behaviour online during extreme events, such as the COVID-19 crisis, to express their awareness, actions and concerns. Hong Kong has implemented stringent public health and social measures (PHSMs) to curb COVID-19 epidemic waves since the first COVID-19 case was confirmed on 22 January 2020. People are likely to engage in collective behaviour online during extreme events, such as the COVID-19 crisis, to express their awareness, actions and concerns. Here, we offer a framework to evaluate interactions among individuals' emotions, perception, and online behaviours in Hong Kong during the first two waves (February to June 2020) and found a strong correlation between online behaviours of Google search and the real-time reproduction numbers. To validate the model output of risk perception, we conducted 10 rounds of cross-sectional telephone surveys from February 1 through June 20 in 2020 to quantify risk perception levels over time. Compared with the survey results, the estimates of the risk perception of individuals using our network-based mechanistic model capture 80% of the trend of people's risk perception (individuals who worried about being infected) during the studied period. We may need to reinvigorate the public by engaging people as part of the solution to live their lives with reduced risk.

**Main Text**

**INTRODUCTION**

Countries have adopted public health and social measures to control transmission, loss of jobs, education and other critical social and cultural activities [1–3]. During extreme events (e.g., COVID-19 in January 2020), people are likely to engage in a miscellaneous set of behaviours (henceforth *collective behaviour*), such as exchanging information in social situations [4]. For instance, people are inclined to swiftly spread messages via social media about natural disasters, in order to gain more knowledge and decrease unforeseen worries [5]. Recent studies suggest people would likely share COVID-19 content on their social media, expressing their awareness, actions and concerns [6,7].

The first COVID-19 case was confirmed in Hong Kong on 22 January 2020 [8]. Since then, Hong Kong has put in place strong measures to prevent COVID-19, including wearing masks in all public areas, closure of schools, bars and social venues, work at home policies, and restaurant measures [9]. Social media has become an all-embracing part of daily life for rapid knowledge dissemination during the COVID-19 pandemic isolation, used by Hong Kong people to express their emotions (e.g., depression) during the COVID-19 pandemic [10]. To study human collective behaviours on covid-19 response, we evaluate interactions among Hong Kong residents' emotions, perception, and behaviours using a network-based mechanistic model that links external situations of COVID-19 prevalence and social networks (**Figure 1A**).

## METHODS

### Data

**Epidemic data**: We collect the daily data of all newly confirmed coronavirus cases by reporting date in Hong Kong from January 31 to June 28, 2020 [11] to denote the impact of external situations to HK residents.

**Google search data**: We collect the daily data of search interest for the disease of Coronavirus disease 2019 in Hong Kong through January 31 to June 28, 2020 from Google Trend [12] to denote the online behaviour dynamics of HK residents.

**Survey data**: In each monthly/weekly survey from February 1 to June 20 in 2020, we contacted either 500 or 1000 local residents through random digit dialing of landlines and mobile telephones, using age, gender, education, and employment information to weight the response frequencies to the adult population in Hong Kong [13]. 8593 local residents were interviewed through these 10 cross-sectional telephone surveys. We asked each participant about the perception of the risk of being infected. Specifically, to assess the risk perception, we asked whether the participant was worried about being infected with COVID-19 (moderately/ very much worried/ extremely worried).

**Modeling collective behaviours during extreme events**

Our stochastic network-based individual model incorporates environment, agents, and local behaviours and updating rules by combining the mapping between multiagent systems and social networks [14]. Following external situation reports, individuals perceive risks, experience differing emotional reactions, and further affect their behaviors following the strengthening process. Inversely, resulting behaviors would reduce individuals' emotion (e.g, anxiety and stress) and risk perception, with respect to the weakening process. More details of methods and data can be found in Supplement. We analyzed the model with a focus on Hong Kong, but the results could be applied to other cities in general.

**RESULTS**

Following reports of external situations, individuals perceive modified risks, experience differing emotional reactions, and further affect their behaviours. Those resulting protective behaviours inversely reduce people's emotion of anxiety and perceptions of susceptibility, resulting in less decreases of perceived risk during March and April 2020 (**Figure 1 C and D**). We projected the daily time series of the observed behaviours of individuals by tracking external situation reports of new infections (**Figure 1 B, C and D**). Comparing observed behaviours of Google search data, our human collective behaviour model incorporating the interactions among agents is well fitted. 79% the observed Google search data are included in the 95% credibility interval (CrI) ranging of normalized averaged levels of individual behaviours across all agents in the model (**Figure 1D**). The Pearson's linear correlation coefficient between the real-time reproduction number ($R_t$) [15] and the averaged levels of individual behaviours is -0.40 with a p-value of 0.0001. After the individual behaviour reached its bottom, $R_t$ began to increase after Apr. 23, 2020.

To validate the model output of risk perception, we conducted 10 rounds of cross-sectional telephone surveys from February 1 through June 20 in 2020. A total of more than 7500 local adult residents have been interviewed via these surveys (see Methods). Such large-scale longitudinal data provides an opportunity to quantify risk perception levels over time. We estimate the daily time series of the risk perception of individuals using our network-based mechanistic model informed by external situation reports of new infections. Compared with the survey results, the estimated values of median and 95% CrI capture 80% (8 out of 10 surveys) of the trend of people's risk perception (individuals who worried about being infected) during the studied period (**Figure 1C**). The average level of individual risk perception continues to decrease, but slows down as case numbers start to soar in March 2020.

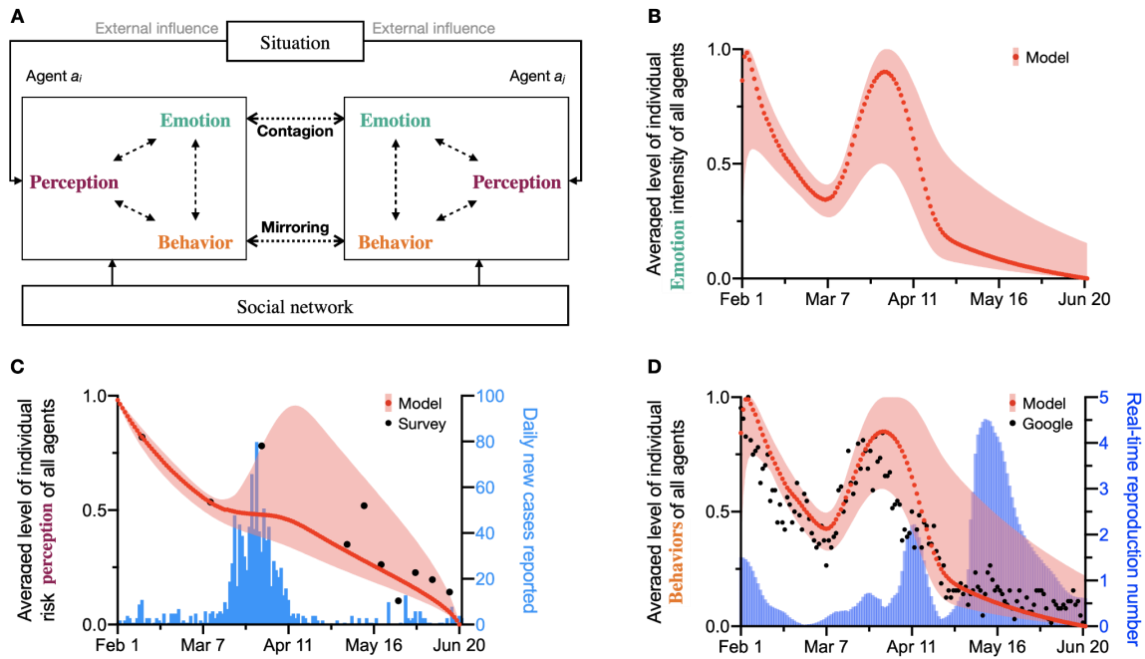

**Figure 1. Reconstruction of human collective behaviours during the first and second waves of the COVID-19 pandemic in Hong Kong from February 1 to June 20 in 2020, using a human collective behaviour model that incorporates daily external situation reports of new infections.** We projected the daily time series of the observed behaviours of individuals by tracking external situation reports of new infections. **(A) Structure of the individual-based model with influences from the external situation and individual-based social networks.** In response to the external situation reports, each individual can experience the strengthening process that first changes the perceived risk of infections and then changes the emotional reactions (e.g, anxiety and stress), which in turn leads to the adjustment of protective behaviours. Each individual will also experience the weakening process, in which the changes in protective behaviours can reduce the emotional reactions and perceived risk of infections. Due to daily interactions on social networks (e.g., messaging in Facebook), the emotion and behaviours of an individual can influence the emotions and behaviours of other connected individuals in the network, denoted as emotion

contagion and behavior mirroring. These dotted lines denote the interactions of psychological factors within and between individuals. **(B) Results for daily individual emotional intensity levels.** The red dots and shading indicate the median and 95% credibility interval (CrI) of square root of normalized averaged levels of individual emotional intensity across all agents in the model. **(C) Results for daily individual risk perception levels.** Our human collective behaviour model is well matched to the observed time series in our survey data (black dots). The black dots indicate the normalized daily logarithmic percentage of individuals who worried about being infected in our survey to indicate people's risk perception in Hong Kong. The red dots and shading indicate the median and 95% credibility interval (CrI) of normalized averaged levels of individual risk perception across all agents in the model. The blue bars indicate daily new cases reported in Hong Kong. **(D) Results for normalized daily search behaviour levels.** Our human collective behaviour model incorporating the interactions among agents is well fitted to the observed time series data (black dots). The black dots indicate the observation of normalized daily Google search behaviours of residents in Hong Kong. The red dots and shading indicate the median and 95% credibility interval (CrI) of averaged levels of individual behaviours across all agents in the model.

## DISCUSSION

Hong Kong has been following a zero-COVID strategy, who has implemented stringent social distancing measures, including unprecedented movement restrictions and quarantines on inbound travelers, universal masking, closure of schools, bars and social venues, work at home policies, and restaurant measures, to curb COVID-19 transmission since January 2020 and bring case numbers down to low levels in each wave [9,16,17]. Given the continued threat of COVID-19 in Hong Kong, pandemic fatigue is a natural response due to complex interplay of cultural and social factors (e.g., the risk perception of threats) [18–20], which has been observed in many countries [21–24]. Our results of the decreasing individual risk perception indicate that the gradual emergence of pandemic fatigue in Hong Kong as demotivation of related behaviours.

Our study has several limitations. First, although we analyzed self-reported behaviour and did not validate this against actual behaviours, self-reported surveys have been widely used to study human behaviour such as contact patterns [25] and hospital attendance [26].

Second, we focus on the period of the first two waves in Hong Kong, which are taken as extreme events, rather than subsequent waves. Informed by the daily Google search interest in Google Trend [12] for COVID-19 in Hong Kong, we find it has decreased 20% to 50% on average since the third wave, perhaps due to people having gained enough knowledge and got used to the COVID-19 situation. Third, other social activities may affect the risk perception and protective behaviours. Prolonged financial stress due to job loss, mask costs and the distrust of government's policies may also contribute to the emergence of pandemic fatigue in the studied period. Despite these limitations, the close matching of model output with human collective behaviour of Google search data and our surveyed risk perception data suggests the capacity of our individual-based human collective behaviour model in capturing the actual population behaviours.

The current socio-political and economic dilemma caused by a pandemic calls for decision-makers to focus beyond the number of cases reported. The fluctuation of human collective behaviours online reflects people's social, emotional and mental health needs, impacted by external situations. To maintain people's risk perception to COVID-19 on a high level, we may need to reinvigorate the public by engaging people as part of the solution, understanding their needs, acknowledging their hardship, and empowering them to live their lives with reduced risk [21,27].

**Acknowledgments:** We acknowledge the financial support from AIR@InnoHK administered by Innovation and Technology Commission of the Research Grants Council of the Hong Kong SAR Government, the Health and Medical Research Fund, Food and Health Bureau, Government of the Hong Kong Special Administrative Region (grant no. COVID190118, 21200632), the Research Grants Council of the Hong Kong Special Administrative Region, China (GRF 17110221), National Natural Science Foundation of China (grant no. 72104208). The weekly telephone surveys were supported by the Government of the Hong Kong Special Administrative Region. BJC consults for AstraZeneca, Fosun Pharma, GlaxoSmithKline, Moderna, Pfizer, Roche and Sanofi Pasteur. BJC is supported by the AIR@innoHK program of the Innovation and Technology